\documentclass[]{acmart}
\renewcommand\footnotetextcopyrightpermission[1]{} 
\usepackage{booktabs} 
\usepackage[]{todonotes}
\usepackage{cleveref}
\usepackage{framed}
\usepackage{color}
\usepackage{tabularx}
\usepackage{multirow}
\newcommand{\etal}{~\textit{et al.}}
\newcommand{\say}[2]{
  \begin{small}
   \textit{``{#1}'' [{#2}]}. 
  \end{small}
}
\newcommand{\theme}[2]{\noindent  \textbf{{{#1}\@. #2}.}}
 
\begin{document}

\title{Needs and Challenges for a Platform to Support Large-scale Requirements Engineering}
\subtitle{a Multiple-case Study}
\author{Davide Fucci}
\orcid{0000-0002-0679-4361}
\affiliation{%
  \institution{HITeC, University of Hamburg}
  \city{Hamburg}
  \state{Germany}
}
\email{fucci@informatik.uni-hamburg.de}

\author{Cristina Palomares, Dolors Costal, Xavier Franch}
\affiliation{%
  \institution{Universitat Politecnica de Catalunya}
  \city{Barcelona}
  \state{Spain}
}
\email{cpalomares,franch,costal@essi.upc.edu}

\author{Mikko Raatikainen}
\affiliation{%
  \institution{University of Helsinki}
  \city{Helsinki}
  \state{Finland}
}
\email{mikko.raatikainen@helsinki.fi}

\author{Martin Stettinger}
\affiliation{%
  \institution{IST, Graz University of Technology}
  \city{Graz}
  \state{Austria}
}
\email{martin.stettinger@ist.tugraz.at}

\author{Zijad Kurtanovic}
\affiliation{%
  \institution{University of Hamburg}
  \city{Hamburg}
  \state{Germany}
}
\email{kurtanovic@informatik.uni-hamburg.de}

\author{Tero Kojo, Lars Koenig}
\affiliation{%
  \institution{The Qt Company}
  \city{Espoo}
  \state{Finland}
}
\email{tero.kojo, lars.koenig@qt.io}

\author{Andreas Falkner, Gottfried Schenner}
\affiliation{%
  \institution{Siemens}
  \city{Vienna}
  \state{Austria}
}
\email{andreas.a.falkner,gottfried.schenner@siemens.com}


\author{Fabrizio Brasca}
\affiliation{%
  \institution{WindTre Italia}
  \city{Milan}
  \state{Italy}
}
\email{fabriziogabrio.brasca@windtre.it}

\author{Tomi M\"annist\"o}
\affiliation{%
  \institution{University of Helsinki}
  \city{Helsinki}
  \state{Finland}
}
\email{tomi.mannisto@cs.helsinki.fi}

\author{Alexander Felfernig}
\affiliation{%
  \institution{IST, Graz University of Technology}
  \city{Graz}
  \state{Austria}
}
\email{alexander.felfernig@ist.tugraz.at}

\author{Walid Maalej}
\affiliation{%
  \institution{University of Hamburg}
  \city{Hamburg}
  \postcode{20257}
}
\email{maalej@informatik.uni-hamburg.de}

\renewcommand{\shortauthors}{D. Fucci et al.}

\begin{abstract}
  \textit{Background}: Requirement engineering is often considered a critical activity in system development projects. 
  The increasing complexity of software as well as number and heterogeneity of stakeholders motivate the development of methods and tools for improving large-scale requirement engineering. 
  \textit{Aims}: The empirical study presented in this paper aim to identify and understand the characteristics and challenges of a platform, as desired by experts, to support requirement engineering for individual stakeholders, based on the current pain-points of their organizations when dealing with a large number requirements. 
  \textit{Method}: We conducted a multiple case study with three companies in different domains. We collected data through ten semi-structured interviews with experts from these companies. \textit{Results}: The main pain-point for stakeholders is handling the vast amount of data from different sources. 
  The foreseen platform should leverage such data to manage changes in requirements according to customers' and users' preferences. 
  It should also offer stakeholders an estimation of how long a requirements engineering task will take to complete, along with an easier requirements dependency identification and requirements reuse strategy. 
  \textit{Conclusions}: The findings provide empirical evidence about how practitioners  wish to improve their requirement engineering processes and tools. 
  The insights are a starting point for in-depth investigations into the problems and solutions presented.
  Practitioners can use the results to improve existing or design new practices and tools. 
\end{abstract}

%
%



\maketitle

\section{Introduction}
Requirements are often considered the basis for all subsequent development, deployment, and maintenance activities. 
Poorly implemented Requirements Engineering (RE) presents significant risks for a project~\cite{DC06}, including its cancellation or additional costs~\cite{HHL01}.
Gartner research found that requirements are the third source of product defects and the first source of delivered defects for service projects. 
Accordingly, the cost of fixing defects ranges from \$70 at the requirements phase to \$14.000 in production phase\footnote{https://www.gartner.com/doc/1753116/hype-cycle-application-development-} .

Nevertheless, RE often receives little project effort~\cite{Fir04} and, in spite of the advances in the field, practitioners still struggle with it~\cite{FFO17, MLM18}.
For instance, a recent study~\cite{FWK17} reports that the requirements definition is a challenge for practitioners. 

At the same time, the latest advancements in machine learning and natural language processing bear new potentials to support decision-makers in the context of RE~\cite{JSB17, MNJ16}. 
As organizations are constantly looking for ways to produce novel products and services, a platform---embedding the above technologies to support the RE process---can  increase stakeholders' and customers' satisfaction, improve time to market, and reduce costs.    

The research project \textsc{OpenReq}\footnote{https://cordis.europa.eu/project/rcn/206364\_en.html} aims to develop such a platform for companies dealing with large-scale requirements---e.g., in the order of thousands per project~\cite{RSW08}.
In particular, \textsc{OpenReq} covers industial scenarios related to bid management for railways, community-driven cross-platform development tools, and telecommunication services.
At the start of the project, it was necessary to clarify the industry needs regarding the lack of support for RE tasks. 
Moreover, the close partnership with industry presented a chance to reduce the divide between practitioners and researchers~\cite{FWK17}.
In fact, as requirements are usually project-specific and business critical,
it is challenging for practitioners to share insights regarding their processes and artifacts.
Accordingly, for researchers it is difficult to form a clear understanding of RE in practice~\cite{FWL12}.   
Understanding the state-of-practice is the first step to support the development of the \textsc{OpenReq} platform in a problem-driven fashion.
As recent studies~\cite{FWK17,FW15} report that stakeholders are overwhelmed by the number of decisions to make when dealing with requirements, we focus our investigation and initial project efforts on the area of \textit{individual stakeholder support}~\cite{PFF18}.

This paper grounds in empirical evidence the desired characteristics for a platform  supporting large-scale RE in industrial context. 
We base our findings on the pain-points associated with how RE tasks are currently approached by three companies in different domains. 
Moreover, we aim at identifying challenges that the companies anticipate when using a platform such as \textsc{OpenReq}. 
To that end, we conducted an exploratory, multiple case study~\cite{Yin17} within the \textsc{OpenReq} partner company units responsible for RE.\footnote{Due to confidentiality agreement the full data (e.g., interviews transcript) cannot be disclosed.} 

The paper makes two contributions. 
First, it offers in-depth insights into the state-of-practice of very-large RE  
and defines the needs for individual stakeholders. 
Second, it generates new insights useful to design a platform supporting such needs, and the potential challenges to be taken into account when adopting it. 

\textbf{Paper organization}.
\Cref{sec:background} surveys the state-of-practice related to current problems and needs in RE. 
\Cref{sec:method} introduces the exploratory case study including research questions, method, and data. 
We present the results and summarize the main findings in \Cref{sec:results}.
We discuss the implication as well as the limitation in \Cref{sec:discussion}.
Finally, \Cref{sec:conclusion} concludes the paper.   

\section{Related Work}
\label{sec:background}
Currently, there seems to be a chasm between the state-of-the-art research and practical RE work~\cite{FWK17}. 
Although there are several empirical studies dealing with the challenges faced by industry during the RE stage, only a small part investigates what industry expects from RE research---e.g., concerning tools and approaches. 

Several studies show that current tool support for RE is not enough. 
Carrillo de Gea\etal~\cite{Car12} presents the results of a survey administered to 38 RE tool vendors with the purpose of gaining insights into tools capabilities and the extent to which tools support RE processes.
Results show that RE tools need improvement, principally regarding requirements modelling, open data model (i.e., available data at any time from an external tool) and data integration features; in particular, the integration of data from external sources. 
A survey of 307 practitioners by Maalej\etal~\cite{MKF14} shows that tool support for complex RE tasks is often poor or absent, with requirements negotiation and planning being the least supported.
In contrast, this study takes the perspective of the tool final users, suggesting what are features relevant for them given the problems they are experiencing in their domain.

Requirements traceability is the main focus of the semi-structured interview study conducted by Rempel\etal~\cite{Rem13}. 
They conducted 20 interviews in 17 companies to investigate how requirements traceability is done by practitioners and the challenges that they face. 
Findings suggest that a traceability strategy defined up front is indeed required because it is a complex task to determine all suitable trace paths for a project. 
The decision for or against trace paths between requirements and artifacts requires a detailed understanding of the project engineering process and goals. 

Palomares\etal~\cite{Pal17} conducted an exploratory survey to investigate the state-of-practice in the reuse of requirements. 
The survey was based on an Internet questionnaire with 71 responses from requirements engineers with industrial experience. 
Although they found that the majority of respondents declared some level of reuse in their projects, only a minority of them declared such reuse as a regular practice.
Ignorance of reuse techniques and processes is the main reason preventing wider adoption, followed by needing a high initial investment, and the complexity of implementing requirements reuse.

Raatikainen\etal~\cite{Raa11} describes the state-of-practice of RE in the nuclear energy domain in Finland on the basis of a descriptive case study focusing on safety automation systems of nuclear power plants. 
Data was collected by interviewing two domain experts representing public authority and five experts working at three power companies. 
The identified challenges of RE are in aging, knowledge transfer, traceability, communication, and tool support. 
As for RE tools, results highlight the need of a tool that supports relationships, hierarchies and traceability, and the difficulty to incorporate them into an existing environment. 

The work of Wnuk\etal~\cite{WRB11} presents the results of 13 interviews in three companies with the objective of understanding the challenges in scoping large RE projects and how these challenges are addressed by the studied companies.
The identified challenges are related to information overload (e.g., difficult to make relevant grouping, hard to have a global overview of the system, how to get deep knowledge, how to acquire the knowledge to make a decision, information cannot be trusted), requirements reuse (inability to reuse requirements), and requirements quality (unclear and complicated requirements). 

Sethia\etal~\cite{Set14} used an online survey to establish and validate an empirical model about the behavior between requirements elicitation issues and project performance. 
The study takes into account the responses of 203 participants involved in RE from different companies and domains. 
The results identify different challenges related to the elicitation of requirements. 
First, there are challenges related to the diversity of stakeholders---e.g., large differences in users' needs, effort needed for reconciling requirements from various users. 
Second, there are challenges related to communication (e.g., users had significant problems when communicating requirements). 
Third, there are problems requirements volatility (i.e., related to the extent of changes that the requirements undergo during the project life cycle). 
Finally, there are problems related to requirements quality, specially concerning requirements that are ambiguous and unstable.
This study confirms the issues identified in~\cite{Rem13,Pal17,Raa11, WRB11,Set14} while elaborating on how to address them in three different domains.

One of the goals of Hiisil\"{a}\etal~\cite{Hii15} is to investigate what are the challenges in the RE process of a customer organization in an outsourced development environment.
With that purpose, they conduct a case study in a Finnish insurance company performing 17 interviews and analyzing 15 large projects. 
In addition, they conducted five workshops to validate the results. 
The case study identifies some challenges related to the elicitation and specification of requirements. 
Some of them are related to information overload (e.g., scoping and planning the project, and prioritization of requirements), some to stakeholders (e.g., large number of stakeholders, lack of cooperation in the early RE stages, management involvement as a stakeholder), some to traceability (from the business requirements to the benefits received), and some to requirements quality (arriving to a common understanding and reaching comprehensive requirements for all parts involved in the RE process). 

To identify key industry needs, Sikora\etal~\cite{Sik11} conducted an in-depth study with representatives from large, internationally-operating companies in the domain of embedded systems in Germany. 
They interviewed ten practitioners with a clear view of the RE needs of their companies.
The authors show that support is needed for traceability, testability, completeness, and verifiability.

On top of recognizing the same problems reported in~\cite{Hii15,Sik11}, this paper presents the needs that experts, working in companies in three different domains and  countries, perceive and the challenges anticipated when addressing them.

Mend\'{e}z\etal~present \textsc{NaPIRE}, a globally distributed family of surveys to study the state-of-the-practice in RE, as well as the results of the first run of the survey in Germany~\cite{FW15} and a second run conducted all over the world~\cite{FWK17}. 
Different challenges related to RE are identified, covering aspects such as missing dependencies, insufficient stakeholder involvement, flaws when communicating with customers, different requirement quality challenges, and the lack of time dedicated to RE activities.

Among the reviewed studies, M\'{e}ndez\etal~\cite{FWK17} is the only one covering the RE needs of companies in several domains and geographical areas. We complement this study by providing qualitative evidence supporting the existence of challenges while focusing on companies dealing with large-scale requirements.

\section{Case Study Design}
\label{sec:method}
\begin{figure*}[t]
    \centering    \includegraphics[width=\textwidth]{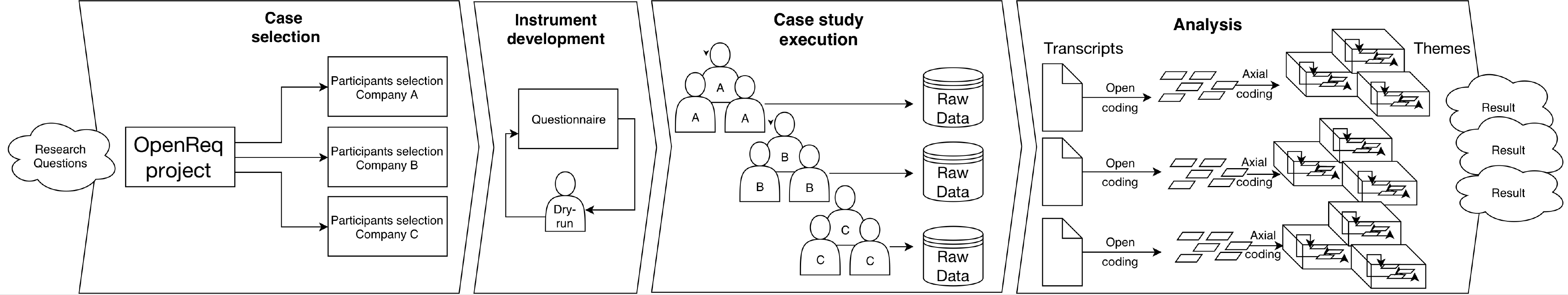}
    \caption{Research phases for the presented case study.}
    \label{fig:method}
\end{figure*}

The objective of this study is to \textit{understand what companies consider an innovative RE platform.} 
Although the interviews covered a broader spectrum of aspects, in this paper, we focus on techniques that support individual stakeholders, meaning techniques that help them, as individuals, during the RE process. These have to
be considered in contrast to the techniques that consider groups of stakeholders in
RE, which takes into account the preferences and needs of all the stakeholders in the group.
We consider this a multiple (i.e., performed through several cases), holistic (i.e., a single unit of analysis is studied in each case~\cite{Yin17}), and exploratory (i.e., the goal is to seek new insights rather than test existing assumptions) case study, reported according to the guidelines of Runeson\etal~\cite{RH09}. 
Data were directly collected using semi-structured interviews with ten company representatives and analyzed using open, axial, and selective coding.
The research process is reported in~\Cref{fig:method}.
\subsection{Research Questions}
Our case study was driven by the following research questions: 
\begin{itemize}
\item\textit{RQ1:} What are the \textit{pain-points} faced by the companies in the way they currently deal with large-scale RE? 
\item\textit{RQ2:} What are the \textit{needs} that a platform supporting large-scale RE should address?
\item\textit{RQ3:} What are the \textit{challenges} of introducing such a platform in the case companies?
\end{itemize}
RQ1 is necessary to get an understanding of the companies status quo and their current problems. 
Thereafter, we investigate the main objective of this research---what needs should the platform support---in RQ2.
Finally, we formulate RQ3 to understand what are the challenges that the companies foresee with the introduction of approaches addressing their needs.    

\subsection{Case and Subject Selection}
All the three companies deal with a large number of requirements; however, several other factors vary between them as summarized in~\Cref{tbl:companies}. 
\Cref{tbl:participants} provides details about the interview participants.
\begin{table*}    
    \centering
    \caption{Overview of the case companies.}
    \label{tbl:companies}
    \small
    \begin{tabularx}{\textwidth}{p{3.2cm} X X X}  
    \textbf{} & \textbf{Company A} & \textbf{Company B} & \textbf{Company C}  \\ \hline
    \textbf{Type of company}   & Software product  & Infrastructure development   & Infrastructure development  \\ \hline
    \textbf{\textsc{OpenReq} scenario} & Cross-platform development tools & Bid management for railways & Telecommunication services \\ \hline
    \textbf{\#employees in RE} & 100 in R\&D, 10 in PM, no specific RE  &   > 200 &  no specific in RE \\ \hline
    \textbf{\#employees in typical project } & All  &  20-30 in different roles &   > 100 \\ \hline
    \textbf{Distribuited}  &  Yes  &  No  &  No  \\ \hline
    \textbf{Source of requirements}                                                            &          Bespoke, community (market-driven)      &            Request for proposal (bespoke)          &       Market driven                 \\ \hline
    \textbf{Main RE tool}  & Atlassian \textsc{Jira} issue tracker  &  IBM Rational \textsc{DOORS}, Microsoft \textsc{Word} &                        \textsc{Remedy}, Microsoft \textsc{Word}, Microsoft \textsc{Excel}\\ \hline
    \textbf{Process model}                                                                     &       Iterative 
    &      Waterfall-like process for the bid phase and iterative or Agile approach for the development phase depending on the specific project                &         Currently being reworked towards Agile              \\ \hline
    \textbf{Release duration for a typical project}                                                                  & Major releases every six months, minor releases even weekly                & For development projects, preparation lasts several weeks to several months  &             Several months           \\ \hline
    \textbf{\#requirements in a typical project}    &      several thousands           &     several thousands  &   hundreds to thousands     \\ \hline
  \end{tabularx}
\end{table*}
\subsubsection{The Qt Company} (henceforth, Company A). 
The company develops tools and solutions for cross-platform software development with a focus on embedded devices. 
The company has several sites worldwide and consists of 200 employees. 
The company is actively involved with open-source development as part of its business is in maintaining and further developing tools and framework freely and openly available to the community. 
Such solutions are then tailored to the needs of the company clients. 
The company has a robust Agile development process in which requirements originate from the community as well from the customers. 
    
The interviewees from Qt are three, one product manager and two senior R\&D managers.
The representative of the \textsc{OpenReq} project in Qt recruited them due to their active involvement in maintaining and improving the RE process within the company. 
The participants worked in the Qt R\&D site in Germany. 
    
\subsubsection{Siemens Mobility AG} (henceforth, Company B).
 The company, part of a larger conglomerate, operates as a supplier and system integrator in the railway sector. The company operates worldwide and has approximately 28,000 employees. Managing bids and requests for proposals (RFPs) for safety systems issued by national providers and ensuring that offers comply with technical specifications, are important tasks of the work process.
One R\&D manager (i.e., the OpenReq representative), one senior R\&D, and two bid managers were interviewed during the study. The participants work at the Austrian location of Siemens.

\subsubsection{WindTre Italia SPA} (henceforth, Company C).
The company is one of the largest telecom operators in Italy and part of a conglomerate which provides services worldwide. 
As it is often the case for companies operating in this sector, their requirements are market driven---i.e., they produce solutions sold to customers on an open-market~\cite{RBB05}.
The unit covered in this study consists of 7,000 employees and has a strong focus on RE activities, while the majority of development is outsourced to third parties.  
    
The interviewees from WindTre are two senior R\&D managers (one of whom is the company representative for \textsc{OpenReq}) and a senior network engineer manager.
The project representative recruited the rest of the interviewees.
They worked in two of the company sites in Italy.
    \begin{table}[h]
        \centering
        \caption{Interviewees. ID indicates the company.}
        \label{tbl:participants}
        \small
        \begin{tabularx}{0.5\textwidth}{p{1cm} p{2.5cm} X X}
        \textbf{ID} & \textbf{Role}            & \textbf{Role \newline experience} & \textbf{Company \newline experience} \\ 
        \hline
        A1 & Product manager &     7 years       &         9 months           \\ \hline
        A2 & Senior R\&D     &      1.5 years       &        13 years            \\ \hline
        A3 & Senior R\&D     &      1.5 years      &        10 years            \\ \hline
        B1 & R\&D manager    &      10 years      &           25 years         \\ \hline
        B2 & Senior R\&D     &   20 years         &  20 years          
        \\ \hline
        B3 & Senior bid manager &    10 years       &     20 years              \\ \hline
        B4 & Senior bid manager &   12 years         &     20 years               \\ \hline
        C1 & Senior R\&D     &    8 years        &        16 years            \\ \hline
        C2 & Senior R\&D     &     20 years      &            22 years       \\ \hline
        C3 & Senior engineer &       18 years     &        18 years            \\ \hline
        \end{tabularx}
        \end{table}
    
\subsection{Data Collection Procedures}
We carried out semi-structured interviews with the participants between March and May 2017. 
For each company, the interviews were co-located and took place at the company premises with all the participants at the same time.
Each interview session lasted for approximately four hours. We took 15 minutes breaks every 45 minutes.
The interviews were carried out in English, and the audio was digitally recorded upon verbal consent from the participants.

The instrument for data collection was a funneled questionnaire~\cite{Sea99}---posing questions from general, based on the research questions, to specific ones.
Therefore, we used open-ended questions to let the participants reflect on improvements, needs, and challenges for their RE process and tools; and closed questions to explicitly obtain information about specific topics (e.g., status-quo, attitudes towards specific solutions). 
After a dry-run, we internally reviewed the questionnaire and adjusted it to fit into the allotted time. 
When doing so, we allowed room for other questions---not included in the script---of potential interest towards our objectives.

A summary of the questionnaire is reported in~\Cref{tbl:questionnaire}, whereas the final interview script is available at the authors' website~\footnote{\url{http://www.upc.edu/gessi/OpenReq/InterviewGuide.pdf}}.
Before the first phase, we presented the objectives of the interview and the case study and explained how the data from the interview will be handled and used. 
\begin{table*}[t]
    \centering
    \caption{Summary of the questionnaire used during the interviews.}
    \label{tbl:questionnaire}
    \small
    \begin{tabular}{llrrl}
    \textbf{Phase} & \textbf{Topics}                                                                                                                                                                                        & \textbf{\# questions} & \textbf{Duration (in minutes)} & \textbf{Research question} \\ \hline
    1     & \begin{tabular}[c]{@{}l@{}}- Background of the company and participants \\ - Status-quo regarding processes and tools for RE\\ - Problems encountered in the current status\end{tabular} & 45                              & 120                                        & RQ1               \\ \hline
    2     & - Ways to improve the current status and addressing pain-points                                                                                                       & 7                              & 50                                        & RQ2               \\ 
    
    2b    & - Improvements specific for individual stakeholders support                                                                                                                                & 8                              & 40                                        & RQ2               \\ \hline
    3     & - Challenges in the implementation of Phase 2                                                                                                               & 5                              & 20                                        & RQ3               \\ 
    3b    & - Challenges  in the implementation of Phase 2b                                                                                                              & 2                              & 10                                        & RQ3     \\ \hline        
    \end{tabular}
    \end{table*}

Up to four interviewees were present during each session; the questions we asked everyone all questions when possible. 
Their answers, comments about others' answers, and conversations were recorded.
When an interviewee in a higher position was present (i.e., product manager), we preferred to hear first the answers from interviewees in other positions (i.e., engineers) to limit biases.  
After the interview, we used a professional service to obtain a timestamped, word-by-word transcript of the audio recording.

\subsection{Analysis Procedures}
We performed qualitative data analysis based on the transcripts obtained as described in~\cite{MH94}. 
We coded the transcripts using a line-by-line approach, in a semi-exploratory fashion. 
The initial codes were based on the main pain-points in RE activities, articipants wish for solutions, and foreseen challenges.
For each transcript relative to a company, each relevant statement in the transcribed interview was assigned codes by two researchers following an open coding approach~\cite{Cha06}.

These codes were subsequently grouped to form themes (e.g., tools, roles, and channels from which requirements arise) and their relationships were captured by applying axial coding~\cite{Cha06}.
The data was analyzed using Atlas.ti\footnote{\url{https://atlasti.com}}, which allows sorting the data according to codes and themes (e.g., by roles, problem, solution) for  easy identification of patterns.
We did not follow a continuous comparison analysis across cases~\cite{Sea99}. 
We coded the three cases independently, with a researcher in charge for a case, and one or two others supporting her with the review of the codes. 
Therefore, common and contrasting themes were later identified for each single case.
This decision was taken because, in the context of the \textsc{OpenReq} project, the solutions will be evaluated in the companies specific contexts.
Therefore, a constant comparison between each case (either during the interviews or coding) could have jeopardized this aspect.\footnote{For the same reason, we do not provide an explicit comparison among the cases and do not explicitly map the codes into RE activities due to their different processes.}  
After the individual case coding, we compared and reconciled the concepts reported in~\Cref{sec:results}.
 
Finally, we applied selective coding to choose the themes relating to individual stakeholders support~\cite{Cha06}.

\subsection{Validity Procedures}
The questionnaire instrument was reviewed by the researchers in two iterations, followed by a complete dry-run to assess possible inconsistency or replicated questions.
The transcript was performed by a third party to avoid introducing researcher bias in the raw data. 
At least two researchers reviewed the transcripts to check its correctness, and gaps were then filled-in by interviewees. 

Triangulation of the analysis was achieved by having different groups of two to three researchers checking the open codes in parallel. 
We maintained traceability between the transcript and the coding, and between the coding and the reporting by using a state-of-the-art software solution for qualitative data analysis which streamlines the process. 

Although the case study lasted for a limited amount of time, the researchers started collaborating with the companies already few months before the case study was planned; some of the authors have long-term cooperations with Company A and Company B.

\section{Results}
\label{sec:results}
This section presents our findings, and answers the research questions.
The final set of codes, reported in~\Cref{tbl:results}, includes the five pain-points the companies felt the most regarding their current support to individual stakeholders in RE, the ten needs that should be addressed, and the 12 challenges that implementing such needs poses. 
In the following, wherever there is a quote from the interviews, there is also between brackets the id of the interviewee who voiced the quote.
\begin{table*}[t]
\centering
    \caption{Themes (pain-points, needs, and challenges) as perceived by the companies.}
    \label{tbl:results}
\small
\begin{tabular}{lccc|lccc|lccc}
\multicolumn{1}{c}{\textbf{\begin{tabular}[c]{@{}c@{}}Pain-point \\ (RQ1)\end{tabular}}} &  A &  B &  C & \multicolumn{1}{c}{\textbf{\begin{tabular}[c]{@{}c@{}}Needs \\ (RQ2)\end{tabular}}} &  A &  B &  C & \multicolumn{1}{c}{\textbf{\begin{tabular}[c]{@{}c@{}}Challenge\\ (RQ3)\end{tabular}}} &  A &  B &  C \\ \hline
\multirow{3}{*}{\begin{tabular}[c]{@{}l@{}}P1. Information\\ overload\end{tabular}} & \multirow{3}{*}{X} & \multirow{3}{*}{X} & \multirow{3}{*}{X} & N1.Data analytics & X & X & X & C1. Data sources & X &  &  \\ \cline{5-12} 
 &  &  &  & N2. Visualization & X &  & X & C2. Heavier process & X &  & X \\ \cline{5-12} 
 &  &  &  & N3. Effort estimation & X & X &  & C3. Lack of metrics & X & X &  \\ \hline
\multirow{3}{*}{\begin{tabular}[c]{@{}l@{}}P2. Tools \\ limitations\end{tabular}} & \multirow{3}{*}{X} & \multirow{3}{*}{X} & \multirow{3}{*}{X} & \begin{tabular}[c]{@{}l@{}}N4. Integration with \\ upstream tools\end{tabular} & X &  & X & \begin{tabular}[c]{@{}l@{}}C4. Changes \\ to process\end{tabular} &  & X & X \\ \cline{5-12} 
 &  &  &  & \begin{tabular}[c]{@{}l@{}}N5. Trace \\ decision-making\end{tabular} & X & X &  & \multirow{2}{*}{\begin{tabular}[c]{@{}l@{}}C5. Community \\ values\end{tabular}} & \multirow{2}{*}{X} & \multirow{2}{*}{} & \multirow{2}{*}{} \\ \cline{5-8}
 &  &  &  & \begin{tabular}[c]{@{}l@{}}N6. Stakeholder \\ feedback\end{tabular} & X & X & X &  &  &  &  \\ \hline
\multirow{2}{*}{\begin{tabular}[c]{@{}l@{}}P3. Dependency \\ between \\ requirements\end{tabular}} & \multirow{2}{*}{X} & \multirow{2}{*}{X} & \multirow{2}{*}{} & \begin{tabular}[c]{@{}l@{}}N7. Configurable\\  dependencies\end{tabular} & X & X &  & \begin{tabular}[c]{@{}l@{}}C6. Indirect\\  dependencies\end{tabular} & X & X &  \\ \cline{5-12} 
 &  &  &  & \begin{tabular}[c]{@{}l@{}}N8. Dependency at \\ different levels\end{tabular} & X & X &  & \begin{tabular}[c]{@{}l@{}}C7. Dependency \\ interpretation\end{tabular} & X & X &  \\ \hline
\multirow{2}{*}{\begin{tabular}[c]{@{}l@{}}P4. Requirements \\ reuse\end{tabular}} & \multicolumn{1}{l}{\multirow{2}{*}{}} & \multirow{2}{*}{X} & \multirow{2}{*}{X} & \multirow{2}{*}{\begin{tabular}[c]{@{}l@{}}N9. Reuse from \\ other domains\end{tabular}} & \multirow{2}{*}{} & \multirow{2}{*}{X} & \multirow{2}{*}{X} & \begin{tabular}[c]{@{}l@{}}C8. Employees \\ turnover\end{tabular} &  &  & X \\ \cline{9-12} 
 & \multicolumn{1}{l}{} &  &  &  &  &  &  & C9. Decay &  &  & X \\ \hline
\multirow{3}{*}{\begin{tabular}[c]{@{}l@{}}P5. Stakeholder \\ identification\end{tabular}} & \multirow{3}{*}{X} & \multirow{3}{*}{X} & \multirow{3}{*}{} & \multirow{3}{*}{\begin{tabular}[c]{@{}l@{}}N10. Personal \\ recommendation\end{tabular}} & \multirow{3}{*}{X} & \multirow{3}{*}{X} & \multirow{3}{*}{} & \begin{tabular}[c]{@{}l@{}}C10. Stakeholder\\  involvement\end{tabular} & X & X &  \\ \cline{9-12} 
 &  &  &  &  &  &  &  & \begin{tabular}[c]{@{}l@{}}C11. Accuracy \\ of the result\end{tabular} &  & X & X \\ \cline{9-12} 
 &  &  &  &  &  &  &  & \begin{tabular}[c]{@{}l@{}}C12. Privacy \\ of the data\end{tabular} & X & X & X \\ \hline
\end{tabular}
\end{table*}
\subsection{RQ1: Pain-points faced in large-scale RE}
\label{sec:rq1}
We report the pain-points which occurred the most, and were shared by at least two of the three case companies.

\theme{P1}{Information overload} 
All the companies have to deal with difficulties due to the size and range of information needed to carry out RE activities.
Stakeholders have to look for relevant pieces of knowledge necessary to make decisions in large, fragmented repositories of data. 
Parsing such information---e.g., for requirements maintenance tasks---is time-consuming and perceived as disheartening:
    \say{There is a certain level of information overload, especially with the mailing list. So I used to read everything, not anymore. I just can't read that anymore, it's just too much under discussion.}{A2}
For example, acquiring data to support requirements elicitation is not a problem, but its analysis is, resulting in diminished productivity: 
    \say{Every project is a bit different but we have all our documents in Sharepoint, and it is hard to find what we need. The problem is that looking at these documents, just to find a single interesting sentence, takes a long time.}{B2}
    
\theme{P2}{Tools limitations}
A second pain-point perceived by the three companies is the limitation imposed by existing tools.

On one hand, Company A main RE tool is too focused on low-level requirements (e.g., bugs to be fixed) and can discourage their community to request larger features or improvements of the products. 
\say{Jira is really driven by bugs. It's lacking when it comes to feature requirements, so it is awkward for people to file more high-level stuff.}{A2} 
On the other hand, Company B uses a tool that  does not offer the possibility to add detailed information to document decisions.
Thus, Company B had to come up with a workaround:
    \say{We fill in the information in DOORS, but then we use hashtags to insert comments, so we can filter it later.}{B4}
However, the interviewees felt that relying on this approach, rather than on tool support, can hinder the traceability between the initial request for proposal and the final requirements.

\theme{P3}{Dependency between requirements}
The stakeholders involved in the companies RE process struggle to identify requirements relationships which can support tasks such as prioritization.

The similarity between requirements is a type of dependency of particular importance for Company B. 
In their case, similar requirements are obtained not only from previous projects but also from other sources, such as stakeholders who are experts in a particular infrastructure: 
\say{There might be similar requirements because the projects are similar. But it can also be that requirements are similar because they are coming from the same stakeholder.}{B1}

Company A leverages dependencies between requirements for prioritization tasks. 
However, understanding the value of managing such dependencies in the long term is a problem: 
    \say{Relationships between requirements gives us sometimes a kind of guidance for prioritization. For instance, when addressing an issue is  required to fulfill another. This dependency raises the question of whether something with no or small benefit should have higher priority than others.}{A1}

\theme{P4}{Requirements reuse}
The problem of identifying dependencies (\textit{P3}), affects the reusability of requirements. 
Company A has basic strategies to reuse requirements from the same sub-domain. 

In Company B, requirements are not reused, but previous knowledge about the project helps the requirements manager to assign a new requirement to a stakeholder.
However, this process is not automated and relies on expert knowledge.
\say{Requirements are not directly reused because they are new and not written by the requirements manager. If there's a requirement that is similar to another bid project, maybe he will remember and look up the person responsible for it.}{B4} 

\theme{P5}{Stakeholders identification} 
The companies follow different strategies to assign the appropriate stakeholder for an RE task.
The company culture has a role in the realization of such strategies.

For example, in Company A, deciding who should be assigned to a requirement is a process driven by open-source values, such as open communication:
    \say{Overrule the initial decision about who should work on something is not a big deal as long as it is discussed together.}{A2}

In the other cases, the decision is driven by the internal company structure and the responsibilities of each department:
    \say{Usually, according to the organization of the company, we know what department is involved in a particular process, so we usually invite these people to start the process [of assigning requirements].}{B1}

Accordingly, the companies perceive that a fixed correspondence between domain and (expert) stakeholder poses a risk related to employees turnover (i.e., low bus factor~\cite{MSW03}).

Moreover, both companies reported a ``domino effect''---a stakeholder who does not believe to be a good fit to process a requirement assigns it to someone else causing delays.

\subsection{RQ2: Needs to be addressed by the plaftorm}
\label{sec:rq2}
In this section, we report the needs the companies reported.
The needs are mapped onto the pain-points presented in~\Cref{sec:rq1}\\
\theme{N1}{Data analytics}
The three companies agreed that leveraging data analytics is necessary to support individual stakeholders in reducing information overload.
In particular, Company A suggested analytics as a way to generate a roadmap to reduce the time stakeholders spend for planning: 
    \say{The use of analytics could get me an understanding of the relative importance of the tasks I need to do---what should I really do next.}{A3}
Company B offered a specific wish for a feature---filtering out information (i.e., from the text) that does not constitute a requirement: 
    \say{A good starting point for us would be differentiating between what is a requirement and what is not.}{B1}

\theme{N2}{Visualization}
To grasp the large amount of data available to them, Company A and Company C identified the need of a visualization---e.g., a dashboard.
In particular, they expressed their need to not only graphically present requirements meta-data (e.g., status, implementation time) but also the decision-making process.
To that end, Company A suggested the use of a heatmap to understand in which phase the stakeholders spend more time.

\theme{N3}{Effort estimation}
The companies need a way to measure the effort necessary to complete a task (e.g., regarding of time) as a way to better plan and limit individual stakeholders from being overloaded.
For Company B, this metric is needed to make deadlines and the associated effort more explicit.
    \say{What we are looking for is to have some indicators, such as <<if you continue at this rate you will not finish in time>> or something similar.}{B2}
At the moment, none of the companies use a data-driven effort estimation approach (e.g.,~\cite{UMB15}) but rely on expert evaluation.

\theme{N4}{Integration with upstream tools}
Company A and Company C perceive as beneficial the integration of their current tools with the ones used to support upstream activities, such as customer satisfaction and sales. 
For example, Company A identified the integration with tools, such as Salesforce, as necessary to get closer to the customers and better identify their needs.
    \say{We would like to see more direct feeding of information from a meeting that a salesperson has with a customer. I know that they have to report how that meeting went, but I don't see it.}{A2}
Company B does not have such need as the requirements originate from, for example, governments with whom interaction is infrequent.

\theme{N5}{Trace decision-making process}
It is essential for Company A and Company B that tools facilitate the traceability of their decision. 
In Company B such possibility is missing. 

\theme{N6}{Stakeholder feedback}
The possibility to give feedback to stakeholders is deemed as a pivotal need by the three companies.
However, such feature is not present in the tools currently in use. 
The companies not only expressed the need for explaining the results to the stakeholder but also to incorporate the stakeholder feedback as part of the results.

\theme{N7}{Configurable dependencies}
Although the companies have in place mechanisms to identify dependencies between requirements, they acknowledge that a ``one-size-fits-all'' definition does not exist.
Accordingly, there is a need for a configurable way of expressing dependencies between requirements.

In the case of Company B, the similarity between requirements is seen as a type of dependency (i.e., duplication of some functionalities) which should be defined differently by each  stakeholder:  
    \say{The term similarity can be seen differently. So, one should configure his measure of similarity because we don't think that we can come up with a fix set.}{B3}

In Company A, dependencies should be identified manually, posing a burden on the users---e.g., when filing a request after consulting the issue tracker. 
For a complex product, this is not feasible, and they recognize the need to automate the process as much as possible.

\theme{N8}{Dependency at different levels}
Dependencies between requirements should be identifiable at different levels---not only within the same requirement document or project but also across projects: 
    \say{Viewing similar requirements either in a project or even between projects, but still identifying if two requirements have been talking about the same thing.}{B1}
In particular, the companies are interested in identifying contradicting requirements between projects. 

\theme{N9}{Reuse from other domains}
The only need that Company B and Company C associate with \textit{P4}, is to base their reuse strategies on other, related domains in which the company has expertise.

\theme{N10}{Personal recommendations}
The interviewees expressed the need to use a (personal) recommender system to identify stakeholders and the requirements they should tackle.
    \say{[...] supporting me with the assignment of the stakeholders, somehow implies that a system has some knowledge about previous decisions, learned from past requirements.}{B1}
For Company A, this need translates into a support system for the triaging of bugs filed in their issue tracker. 
    \say{What could certainly help is if the system can suggest me an assignee [for a bug].}{A1}

\subsection{RQ3: Challenges in introducing platform}
In this section, we report an overview of what the companies perceive as the main challenges for the implementation of the needs identified in~\Cref{sec:rq2}.

\theme{C1}{Data source}
According to Company A, the data for analytics should be carefully chosen otherwise resulting in more information overload: 
    \say{The problem is that if you start collecting one [type of data], then there are fifty different other things that you want, and then it becomes questionable very quickly.}{A3}
The other two companies did not report this as a challenge due to the predictability of the requirements sources.

\theme{C2}{Heavier process}
Company A perceives that a data analytics solution can hinder the ``agility'' of their process: 
\say{We also have to watch out that that it doesn't make the whole process longer. For example, if the tool pushes you to look through all this other information, you just give up because you just want to file a bug and be done with it.}{A2}

\theme{C3}{Lack of metrics for effort estimation}
Realizing a mechanism to estimate the effort required to complete an RE task necessitates specific metrics, not currently collected within the companies.

\theme{C4}{Changes to the process}
The interviewees expressed their worries regarding tool integration, fearing that it will drive a change in their usual process:
    \say{the other difficulty is that, to follow the new process, it is necessary to compose the requirement in another way.}{B2}

\theme{C5}{Community values}
With \textsc{Jira} integrated with other tools (e.g., CRM) to document decisions taken in Company A, it is necessary to respect the community values when communicating them.
    \say{I'm all for transparency, but this implies that sometimes you have to argue with customers, and explain them why you are downgrading the priority for the delivery of a component.}{A2}
Besides, they fear that customers will not accept decisions exclusively based on the tool results. 

\theme{C6}{Indirect dependencies}
The companies perceived indirect dependencies as a hurdle to address requirement dependencies in general. 
This is particularly challenging when requirements are contradicting:
    \say{Identifying contradicting requirements is an hard task. Perhaps one requirement requires a certain approach and another requires another approach, but two those approaches are not compatible.}{B1}

\theme{C7}{Dependency interpretation}
A challenge, reported by both companies, is the different interpretations that stakeholders attribute to dependencies. 
Giving the possibility to identify dependencies at different levels is risky as stakeholders have to interpret or asses dependencies from unfamiliar projects.

\theme{C8}{Employees turnover}
Company C indicated the change in personnel dealing with requirements as a threat to requirement reuse: 
    \say{Sometime the system where you can reuse a requirement could change. But also the people in charge of the system could change. This is, in my opinion the major difficulty to reuse completely a requirement.}{B3}
The companies perceive as relevant the human factors in reuse.

\theme{C9}{Requirements volatility}
Company C identified requirements volatility---i.e., becoming less relevant over time---as a problem for reuse, especially when changes are exogenous to the company. 
    \say{I think it's hard for a process, or a solution... everything changes very fast. For example, for new regulation, we changed our approach many times}{C1}

\theme{C10}{Stakeholder involvement}
Identifying stakeholders necessitates their modeling (e.g., based on preferences, career profile, attitudes).
Company A and Company B perceived finding and leveraging expert knowledge as a barrier:
    \say{Even if we go through all the old projects, we need an expert... That's the hard part.}{B3}
    
\theme{C11}{Accuracy of the results}
Two of the companies observed that it is not only essential to have a system that helps them during RE, but also that the results for complex tasks (such as recommendations or data analytics) should be as accurate as possible. 
Otherwise, the effort needed for correcting the results will discourage users from using the system.
    \say{For me, this [a tool] is successful if it can analyze a lot of data and give us the same results that a human can give.}{B2}
In other words, as proposed by Berry~\cite{Ber17}, the tool should achieve better recall than a human working on the task manually.

\theme{C12}{Privacy of the data}
All companies agreed that it might be difficult to have access to the data that a RE system to support individual stakeholders might need (e.g., information about users or monitoring the use of the tool). 
This is due to company privacy policies, or due to legal aspects.

\section{Discussion}
\label{sec:discussion}
Our results reflect the opinion of companies operating in different countries, dealing with different domains, and using different process models and tools. 
The results further the current RE community effort to move towards a problem-driven research agenda, where solutions are discussed and developed together with industry~\cite{FW15}. 
In this section, we report the main implications of our work together with its limitations.
\subsection{Implications}
\label{sec:implications}
The implications are based on the results presented in \Cref{sec:results} as well as the state-of-practice reported in \Cref{sec:background}.
We suggest three focus areas for designing a platform to support large-scale RE.

\noindent\textbf{Strategies to fight back information overload}. 
Out of the identified problems, as can be expected for companies dealing with large-scale requirements, the main one is information overload. 
Accordingly, the causes for the other pain-points reported can be traced back to the amount and diversity of data the stakeholders need to deal with, as their tasks (e.g.,  requirements dependency or stakeholder identification) become harder due to the scale of the data. 
However, information overload does not appear among the problems currently perceived in the current RE state-of-practice~\cite{FWK17}.
This observation holds even when only companies operating at a similar scale to the one considered in this study are taken into account.   
On the other hand, fighting information overload is one of the research areas identified by Regnell\etal in the context of large- and very large-scale RE~\cite{RSW08}.

Directly related to this problem is what Wnuk\etal~\cite{WRB11} call ``scoping''---i.e., a set of mechanisms to cope with requirements scalability problems. 
Although not explicitly mentioned during the interviews, ``scoping'' is what the companies would like to achieve through better data analytics solutions. 
Our results showed two specific instances in which data analytics should be applied to cope with information overload, visualization, and effort estimation. 
Visualization is perceived as a way to see the ``big picture''---a need already expressed in~\cite{WRB11}; whereas effort estimation can help to realize a more predictable process when lots of requirements need to be handled.
Although effort estimation is well-studied in software development (in open source~\cite{WPZ07} or otherwise~\cite{UMW14}), few studies address it from an RE perspective despite other industrial studies pointing to its importance (e.g.,~\cite{HCG16}).
In practice, trying to control information overload through effort estimation does not seem to be actively pursued as the primary mechanisms to fulfill such need (i.e., collecting the necessary metrics to calculate effort) are not in place. 
The companies did not explain why that may be the case. 

The fear of complicating the process and the selection of the sources for data analytic approaches are strictly related. 
On the one hand, it is not recommended to add new data which is likely to not bring improvement while impacting the process. 
On the other hand, discarding a data source can introduce a bias in the analytic.
Such decision should be evaluated in the context of the company. 
For example, in the case of Company A, such scenario is sensible due to the level of public accountability of the company towards its community.
We recommend further investigations on information overload and scoping in large-scale RE to focus on the feasibility of effort estimation techniques. 
Moreover, trade-offs---based on the company context---need to be taken into account. 

\noindent\textbf{Stakeholder role in requirements dependency and reuse.}
Both requirements dependency and reuse were identified as pain-points, supporting the  results from previous exploratory studies~\cite{Pal17, Raa11}. 
However, a new factor emerged in our study, the importance of different stakeholders in developing these approaches.
On one hand, not only artifact-based features (e.g., the project to which a requirement belongs to) but also the requirement engineer personal preferences need to be taken into account when defining the dependencies.
On the other hand, attention should be given to the fact that other stakeholders should be able to interpret the dependencies. 

Previous research has pointed out the ``people'' factor as a barrier for the adoption of requirement reuse.
For example, Palomares\etal~\cite{Pal17} reports the resistance to change of requirements engineers as the most prominent of such barriers.
Instead, the challenge we observed is the loss of knowledge, necessary for reuse, due to employees turnover.    
In turn, this observation points to the importance of knowledge maintenance for requirements reuse in line with the results presented in~\cite{Pal17,HJH14} and, in general, supports the recommendation of Dyba~\cite{Dyb05} for software process improvements.
We report volatility of requirements as an additional challenge to reuse. 
This further supports the result of Sethia\etal~\cite{Set14} which identifies such factor as one of the most prominent reasons for project failures.

The results of this study remark the importance of including stakeholders in the definition of requirements dependencies and reuse.
For the former, particular attention should be given to configurability and interpretability; for the latter focus should be given to manage both endogenous (i.e., within the company) and exogenous (i.e., from external factors) changes.

\noindent\textbf{Integration of stakeholder in tool-based decisions.}
From our results, it emerged that RE tools are needed to create a holistic view of the decisions about requirements.
However, to reach such objective, integration should be as easy as possible.
Nevertheless, Camarillo deGea\etal~\cite{Car12} shows that only a few tools support open exchanging formats, such as ReqIF.
This is needed especially for vertical integration, since companies work on different requirements repositories according to their business units and with different tools.
Focus on the ``people'' factor is also necessary for better tooling. 
Maalej\etal~\cite{MKF14} shows that the requirement author intention is one of the top information needs sought by a stakeholder when understanding a requirement, and at the same time, it is one of the least supported features in current tools. 
Accordingly, our study shows the need for stakeholders to incorporate their feedback as part of the results presented by a tool.
Such mechanism can improve the transparency of the decision-making process, thus providing first support for resolving conflicts between stakeholders---i.e., the other information need currently ill-supported by RE tools~\cite{MKF14}.

\subsection{Limitations}
In this section, we briefly summarize the limitations of our case study based on the guidelines presented in~\cite{RH09}.\\
However, as interviews were the only data collection method, we could have missed relevant information.\\
\noindent\textit{Construct validity}. 
This threat can arise once there is not a shared understanding of the terms used in the interviews between interviewers and interviewees.
In our study, the interviews were based on concepts from the domain of the stakeholders who participated in the study. 
The company representatives were involved in the creation of the interview script.
At the beginning of each interview session, we dedicated time to go through the status quo of the company with the participants. 
In addition, some of the interviews were carried out with managers and some of their employees. In our study, we believe this is no big issue since: 1) since the companies where the interviews where this happened do not follow a hierarchical employee structure (i.e., there is no fear to managers), and 2) The employees were really participative in the interviews. However, we cannot control some information was missed out because of that.
Triangulation with other data collection methods (e.g., observation) could have improved construct validity. 
\\
\noindent\textit{Internal validity}. 
This study does not involve the assessment of causal relationships using statistical methods, as the case study is purely descriptive. 
The recommendation presented in~\Cref{sec:implications} are based on our interpretation of the codes, their co-occurrence, and their importance as perceived by the participants. 
The themes emerged after at least two researchers reviewed the codes. \\
\noindent\textit{External validity}. 
As it is usual for case studies, we do not claim strong generalization of our findings. 
We showed that the results apply to different extent within the case companies. 
We make this explicit when reporting the findings by indicating the context in which they are more sensible.
Only some of our results (e.g., the pain-point of information overload) generalize over the three cases. 
For these results, we do not claim statistical generalization to a pre-defined population of companies due to the low number of cases and the non-probabilistic sampling approach driven by the project settings. 
However, they are important in the context of the project to, for example, prioritize what needs should be fulfilled when developing a solution.
We claim analytical generalization of our findings by expanding the theory current on RE-related pains and needs presented in M\'{e}ndez\etal~\cite{FWK17} when considering companies dealing with large-scale requirements in a bespoke or market-driven domain.
\\
\noindent\textit{Reliability}. 
We strengthened the reliability of the interview scripts by running an internal pilot; moreover, we used established coding techniques and tools to code the collected data.
On the other hand, we acknowledge that the researchers, as well as some of the interviewees, had in-depth knowledge on the context in which the case studies were carried out---e.g., the \textsc{OpenReq} project.
Therefore, other researchers wanting to carry-out the same study need to familiarize with the background and objectives of the project. 

\section{Conclusion}
\label{sec:conclusion}
In this paper, we report the results of a multiple case study with three companies dealing with large-scale RE.
The interviews with ten requirement engineers indicate information overload, requirements dependency identification, and current tool limitations as the main pain-points for individual stakeholders. 
Accordingly, a platform to support large-scale RE should address the need of gathering data through existing tools integration to offer analytic solutions.
A sensible RE task for such data pipeline is effort estimation. 
Moreover, we showed that stakeholders preferences must be included when devising approaches for requirements dependency and reuse patterns identification. 
Similarly, the platform should take into account stakeholders feedback and trace decision-making from upstream in the RE process. 
Finally, we identified several challenges for the adoption of such a platform.

We will further validate our findings through focus groups and confirmatory surveys.
Finally, we will evaluate a prototype version of the platform with the companies involved in this study.
\section*{Acknowledgments}
The authors of this paper have received funding from the European Union Horizon 2020 research and innovation Programme under grant agreement No 732463.

\bibliographystyle{ACM-Reference-Format}
\bibliography{esem18_biblio}

\end{document}